\documentclass[aps,prl,twocolumn,groupedaddress,floatfix,superscriptaddress]{revtex4-1}

\usepackage{amsmath}
\usepackage{amssymb}
\usepackage{graphicx}
\usepackage{float}
\usepackage[english]{babel}
\usepackage{dsfont}
\usepackage{epstopdf}
\usepackage{soul}
\usepackage{color}
\usepackage{braket}
\usepackage[colorlinks=true,
            linkcolor=magenta,
            urlcolor=blue,
            citecolor=blue]{hyperref}

\usepackage[normalem]{ulem}
\emergencystretch=\maxdimen
\begin{document}
\title{Experimental observation of Aharonov-Bohm cages in photonic lattices}
\author{Sebabrata~Mukherjee}
\email{mukherjeesebabrata@gmail.com}
\affiliation{Scottish Universities Physics Alliance (SUPA), Institute of Photonics and Quantum Sciences, School of Engineering and Physical Sciences, Heriot-Watt University, Edinburgh EH14 4AS, United Kingdom}
\author{Marco Di Liberto}
\affiliation{Center for Nonlinear Phenomena and Complex Systems, Universit\'e Libre de Bruxelles, CP 231, Campus Plaine, B-1050 Brussels, Belgium}
\author{Patrik \"Ohberg}
\affiliation{Scottish Universities Physics Alliance (SUPA), Institute of Photonics and Quantum Sciences, School of Engineering and Physical Sciences, Heriot-Watt University, Edinburgh EH14 4AS, United Kingdom}
\author{Robert R.~Thomson}
\affiliation{Scottish Universities Physics Alliance (SUPA), Institute of Photonics and Quantum Sciences, School of Engineering and Physical Sciences, Heriot-Watt University, Edinburgh EH14 4AS, United Kingdom}
\author{Nathan~Goldman}
\affiliation{Center for Nonlinear Phenomena and Complex Systems, Universit\'e Libre de Bruxelles, CP 231, Campus Plaine, B-1050 Brussels, Belgium}

\begin{abstract}
We report on the experimental realization of a uniform synthetic magnetic flux and the observation of Aharonov-Bohm cages in photonic lattices. Considering a rhombic array of optical waveguides, we engineer modulation-assisted tunneling processes that effectively produce non-zero magnetic flux per plaquette. This synthetic magnetic field for light can be tuned at will by varying the phase of the modulation. In the regime where half a flux quantum is realized in each plaquette, all the energy bands dramatically collapse into non-dispersive (flat) bands and all eigenstates are completely localized. We demonstrate this Aharonov-Bohm caging by studying the propagation of light in the bulk of the photonic lattice. Besides, we explore the dynamics on the edge of the lattice and discuss how the corresponding edge states can be continuously connected to the topological edge states of the Creutz ladder. Our photonic lattice constitutes an appealing platform where the interplay between engineered gauge fields, frustration, localization and topological properties can be finely studied.
\end{abstract}

\maketitle

{\it Introduction.$-$}
The investigation of electron transport in crystals subjected to external magnetic fields has led to the discovery of many intriguing phenomena including the integer and fractional quantum Hall effects~\cite{Klitzing1980new, Tsui1982two}. In recent years, inspired by the idea of quantum simulation~\cite{Feynman1982simulating}, there has been a growing interest in realizing and exploiting synthetic magnetic fields for electrically neutral particles in artificial crystals, such as cold atoms in optical lattices~\cite{Lin2009synthetic, Aidelsburger2013realization, Miyake2013realizing}. Synthetic magnetic fields have also been realized for photons, thus paving the way to the emergent field of Topological Photonics \cite{Lu2014, Khanikaev2017, Ozawa2018}. For example, a uniform magnetic field has been realized using off-resonantly coupled ring resonators~\cite{Hafezi2013imaging}, and a strain-induced pseudo-magnetic field has been demonstrated in coupled-waveguide arrays~\cite{Rechtsman2013strain}. Additionally, Landau levels physics has been explored in a multimode ring resonator through the realization of a fictitious Lorentz (or Coriolis) force acting on photons~\cite{Schine2016synthetic}.

The quantum properties of a charged particle moving in a magnetic field find their origin in the Aharonov-Bohm (AB) phase~\cite{Aharonov1959significance, Wu1975}, namely, the phase acquired by the wave function of the particle as it performs a loop around a spatial region containing a non-zero magnetic flux. Direct evidence of this AB phase has been experimentally demonstrated in electronic systems~\cite{Chambers1960shift}, and more recently, in photonics~\cite{Satapathy2012classical, Li2014photonic} and ultracold atoms~\cite{Duca2014}. 
Interestingly, for certain geometries and specific values of transverse magnetic fields, noninteracting particles can exhibit {\it complete localization}~\cite{Vidal1998aharonov,Vidal2000interaction} due to destructive interferences of the wavefunction. This AB caging phenomenon, which was found to occur in the $\mathcal{T}_3$ (or dice) and the rhombic lattice, is distinct from Anderson localization~\cite{Anderson1958absence}, which is instead caused by disorder. An experimental signature of this phenomenon was reported in solid-state, by detecting a depression of the critical current and that of the superconducting-transition temperature in a network of superconducting wires~\cite{Abilio1999magnetic}, and also by magnetoresistance measurements in normal metal networks~\cite{Naud2001aharonov}.

\begin{figure}[t!]
\center
\includegraphics[width=8.6 cm]{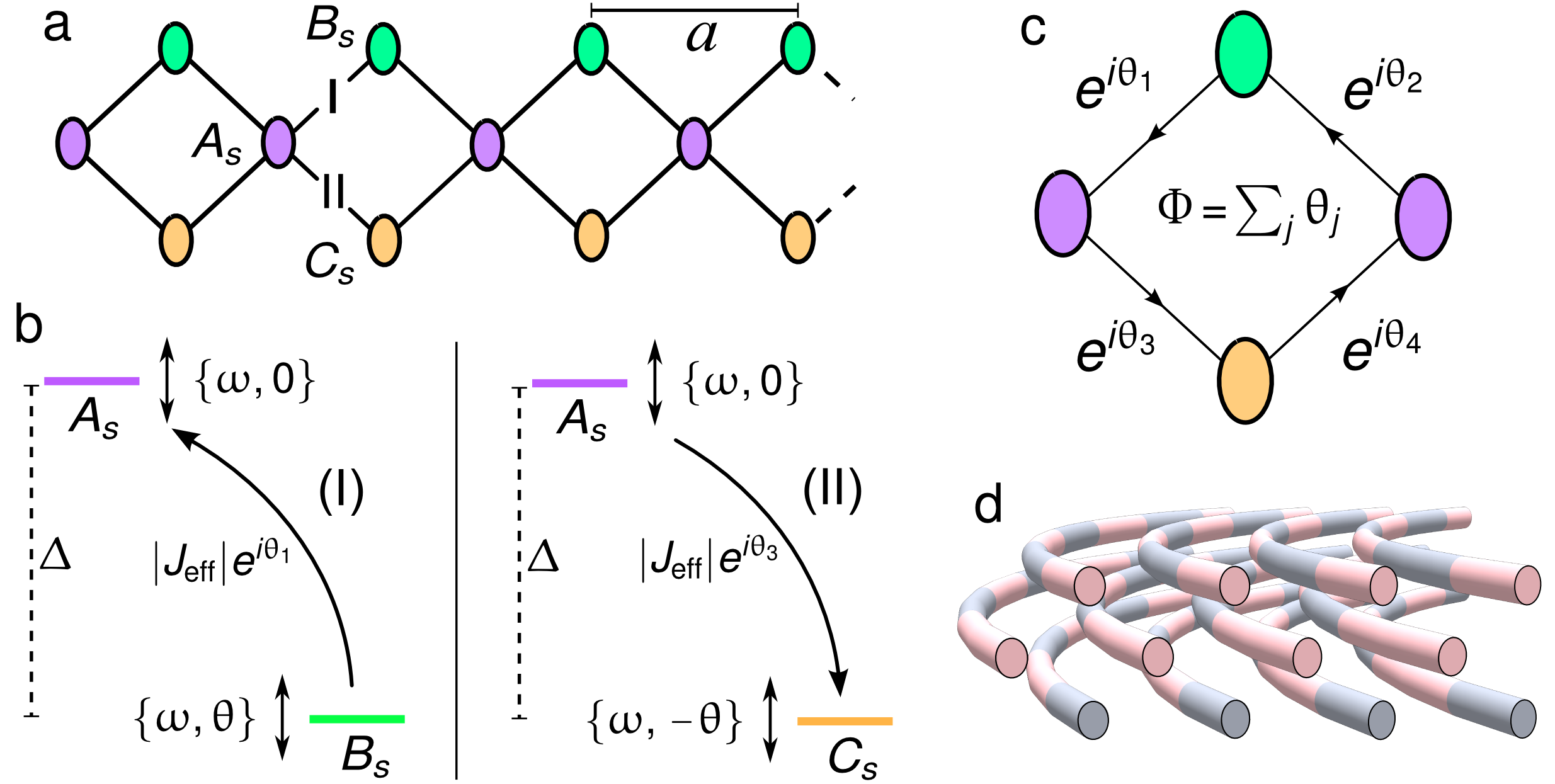}
\caption{(a) Schematic diagram of a rhombic (or diamond) lattice with three sites ($A$, $B$ and $C)$ per unit cell. 
(b) Complex-valued tunneling amplitudes are generated by applying a linear gradient of onsite energy $\Delta$ along the lattice and then periodically modulating the onsite energies with frequency $\omega\!=\!\Delta$ and phase $\theta$, see Fig.~\ref{fig2}(a).
(c) In this driving protocol, a non-zero synthetic magnetic flux ($\Phi$) per plaquette is realized and it can be tuned by varying the phase of modulation $\theta$.
(d) Simplified sketch illustrating the 3D waveguide paths of a driven photonic rhombic lattice with $\Phi\!=\!\pi$ flux per plaquette. The linear gradient of onsite energy is realized by circularly curving the lattice. Here, the color contrast indicates {\it a square wave modulation} of the onsite energy. 
}
\label{fig1}
\end{figure}

Here, we demonstrate the first experimental realization of a uniform synthetic magnetic flux in ultrafast-laser-fabricated waveguide arrays, and use this setting to observe AB cages for light. The idea of creating and controlling a synthetic magnetic flux for photons propagating in a lattice was first proposed in Ref.~\cite{Fang2012realizing}, where a periodic modulation of the inter-site tunneling amplitudes was shown to realize complex effective tunneling matrix elements; see also Refs.~\cite{Longhi2014aharonov,Spracklen2018}. In this work, we follow another approach~\cite{Longhi2014aharonov}, and apply a linear detuning of the propagation constants along the lattice, in order to suppress the effective inter-site tunneling; this tunneling is then restored and controlled by resonantly modulating the propagation constants with a desired phase of modulation. Complex-valued tunneling matrix elements and a non-zero synthetic magnetic flux are successfully generated through this resonant modulation-assisted tunneling process.
We build on this scheme to realize AB cages for photons evolving on a rhombic  geometry [Fig.~\ref{fig1}(a)]. In the caging limit, which is reached when each plaquette is associated with a flux of $\pi$, all the energy bands of the lattice become dispersionless (flat). 
By launching input states that overlap with these flat-bands, we observe  breathing motions of the optical intensity, which clearly signal photonic AB caging.
We then demonstrate that the overall bandwidth of the spectrum can be well-controlled by varying the driving parameters. We also present the dynamics of the states localized on the edge, which can be continuously connected to the topological mid-gap edge states of the Creutz ladder~\cite{Creutz1999,Bermudez2017}.

{\it Driving protocol.$-$} 
Consider a quasi-1D rhombic lattice~\cite{Vidal2000interaction, Creffield2010coherent, Longhi2014aharonov, Khomeriki2016landau, Mukherjee2015rhombic} with three sites ($A$, $B$ and $C)$ per unit cell, Fig.~\ref{fig1}(a). In the tight-binding approximation, the single-particle Hamiltonian with only nearest-neighbor tunneling processes reads 
\begin{eqnarray}
\label{H0}
&\hat{H}_0\!=\!-J \sum\limits_s (\hat{a}_s^{\dagger} \hat{b}_s
+\hat{a}_s^{\dagger} \hat{c}_s
+\hat{a}_s^{\dagger} \hat{b}_{s-1}
+\hat{a}_s^{\dagger} \hat{c}_{s-1}+{\text{H.c.}}), \ \label{H0}
\end{eqnarray}
where $J$ is the tunneling amplitude between neighboring sites, $\hat{a}_s^{\dagger}$, $\hat{b}_s^{\dagger}$ and $\hat{c}_s^{\dagger}$  ($\hat{a}^{}_s$, $\hat{b}^{}_s$ and $\hat{c}^{}_s$) are the creation (annihilation) operators for a particle at the $A$, $B$ and $C$ sites of the $s$-th unit cell, respectively. This lattice supports three energy bands, two of them are dispersive and the third one is perfectly flat at zero energy~\cite{Mukherjee2015rhombic}. 
To realize complex-valued tunneling amplitudes leading to a non-zero magnetic flux per plaquette, the Hamiltonian of the system is engineered in the following way. First, the onsite energies are linearly detuned along the lattice 
\begin{equation}
\hat{H}_{\text{d}} = -\Delta \sum\limits_s \left( 2s\,\hat{a}_s^{\dagger} \hat{a}_s
+(2s+1) \hat{b}_s^{\dagger} \hat{b}_s +(2s+1) \hat{c}_s^{\dagger} \hat{c}_s \right), \label{H_dc}
\end{equation}
where $\Delta$ is the onsite energy shift between nearest-neighbor sites, which is considered to be large, $\Delta \gg J$, to inhibit effective tunneling (an effect known as Wannier-Stark localization~\cite{Mukherjee2015WSL}). The onsite energy offsets are then independently and periodically modulated in time
\begin{equation}
\hat{H}_{\text{ac}}\!=\!\frac{K}{2}\sum\limits_s \left( f(\omega t)\hat{a}_s^{\dagger} \hat{a}_s\!+ \!f(\omega t\!+\!\theta)\hat{b}_s^{\dagger} \hat{b}_s\!+\! f(\omega t\!-\!\theta)\hat{c}_s^{\dagger} \hat{c}_s\right), \label{H_ac}
\end{equation}
where $K$ and $\omega(\equiv 2\pi/T)$ are the peak-to-peak amplitude and frequency of a square wave modulation ($f$), respectively, $+\theta$ ($-\theta$) is the relative phase of modulation between the $A$ and $B$ ($C$) sites.  Considering the total time-dependent Hamiltonian, $\hat{H}(t)\!=\!\hat{H}_{0}+\hat{H}_{\text{d}}+\hat{H}_{\textrm{ac}}$, a resonant ac modulation (i.e.~$\Delta\!=\!\omega\nu$ where $\nu \in \mathbb{Z}$) can restore effective tunneling processes~\cite{Longhi2014aharonov}, as depicted in Fig.~\ref{fig1}(b). The quasienergy spectrum for this periodically driven system is obtained by diagonalizing the Floquet/effective Hamiltonian, $\hat H_{\textrm{eff}} \equiv (i/T)\log\hat U(T)$, where the time-evolution operator over one period is defined as $\hat{U}(T)\!=\!\mathcal{T} \exp(-i\int_0^{T} \hat{H}(t') {\text{d}}t')$, $\mathcal{T}$ indicates the time ordering and $T$ is the period of driving~\cite{Goldman2015,Eckardt2015}.

In the high-frequency limit $\omega \!\gg \!J$ and for a resonant modulation, one can employ standard methods~\cite{Goldman2015,Eckardt2015} to calculate a systematic expansion in $1/\omega$ of the operator $\hat H_{\textrm{eff}}$. The lowest order of this expansion is of the form 
\begin{align}
\hat H_{\textrm{eff}}^{(0)} =|J_{\text{eff}}| \sum\limits_s \big(&e^{i\theta_1}\hat{a}_s^{\dagger} \hat{b}_s
+e^{-i\theta_2}\hat{a}_s^{\dagger} \hat{b}_{s-1} 
+e^{-i\theta_3}\hat{a}_s^{\dagger} \hat{c}_s \nonumber\\
&+e^{i\theta_4}\hat{a}_s^{\dagger} \hat{c}_{s-1}\!+\!{\text{H.c.}}\big), \ \label{Heff}
\end{align}
where the restored tunneling amplitudes are now complex-valued and a non-vanishing constant flux $\Phi\!=\!\sum_j \theta_j$ is realized in each plaquette [Fig.~\ref{fig1}(c)].

\begin{figure}[]
\center
\includegraphics[width=8.6 cm]{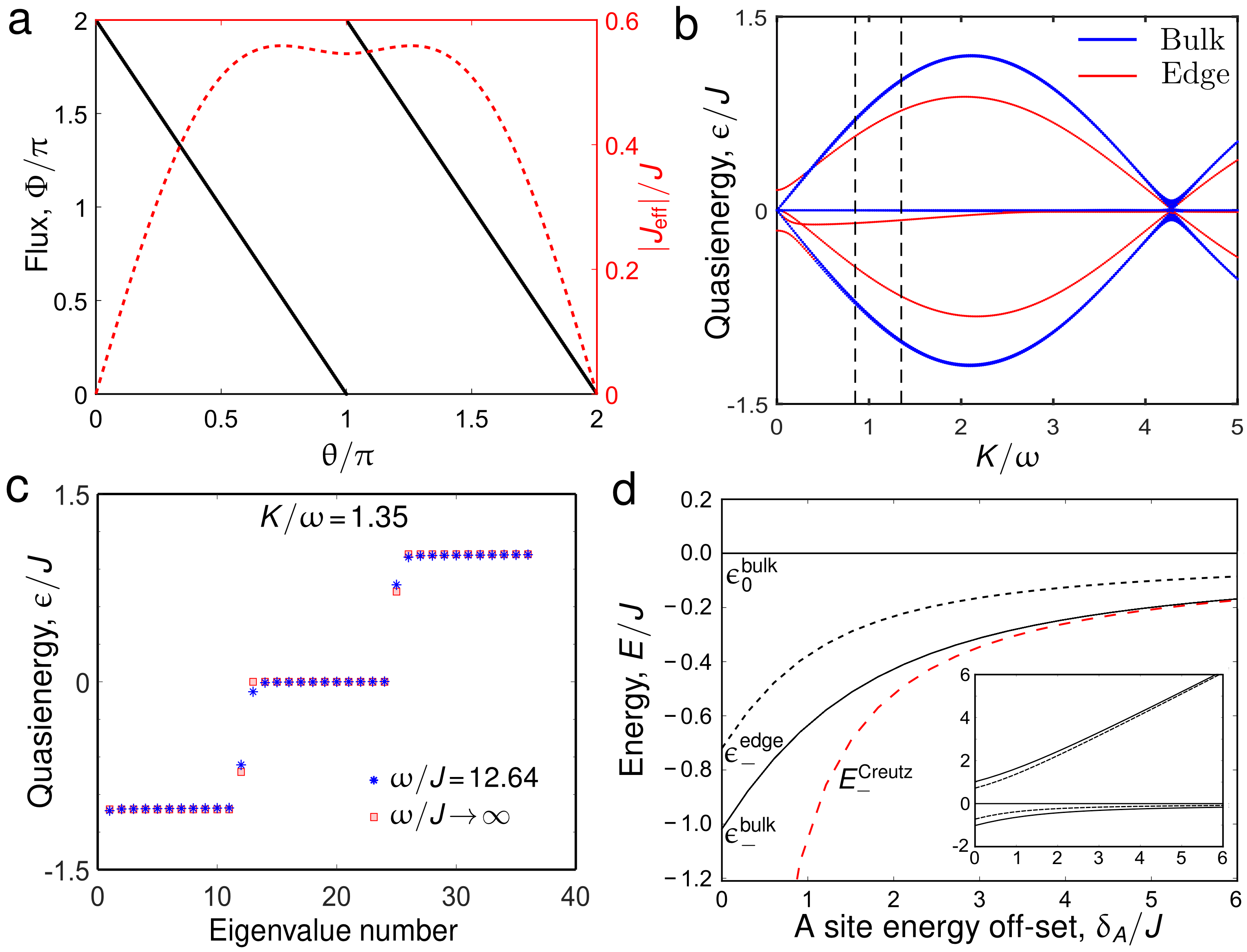}
\caption{(a) Variation of flux and the modulus of effective tunneling amplitude as a function of the phase of modulation ($\theta$) for $K/\omega\!=\!1.35$. 
(b) Quasienergy spectrum for a finite rhombic chain with $\theta\!=\!\pi/2$. Blue and red colors are associated with the bulk and edge modes respectively. The dashed vertical lines indicate the values of $K/\omega$ ($1.35$ and $0.85$) that were realized in the experiment. (c) A comparison of the Floquet quasienergy spectrum for $\omega/J\!=\!12.64$ and $\omega/J\!\rightarrow\!\infty$. Here, $K/\omega\!=\!1.35$. 
(d) Zoom of the energy spectrum of $\hat H^{(0)}_{\textrm{eff}}$ with an additional detuning $\delta_A$ of the onsite energy at $A$ sites (full spectrum shown in the inset). The red dashed line is the energy ($E_{-}^{\text{Creutz}}\!=\!-4J^2_{\textrm{eff}}/\delta_A$) of the lowest flat-band of the Creutz ladder into which we map $\hat H_{\textrm{eff}}+\hat H_{\delta_A}$ for $\delta_A\gg J_{\textrm{eff}}$. The upper flat-band of the Creutz ladder is at energy $E_{+}^{\text{Creutz}}\!=\!0$. The black dotted line corresponds to the edge state, which becomes a topological mid-gap state (protected by an emergent chiral symmetry) in the Creutz-ladder limit, $\delta_A\gg J_{\textrm{eff}}$. Parameters are chosen as in (c).
}
\label{fig2}
\end{figure}

Fig.~\ref{fig2}(a) shows how the synthetic flux $\Phi$ and the modulus of the effective tunneling amplitude $|J_{\textrm{eff}}|$ vary as a function of the phase of modulation $\theta$. The AB caging condition that yields a full localization of the eigenstates is met when $\Phi\!=\!\pi$, i.e.~when the phase of modulation is $\theta\!=\!\pi/2$ or $\theta\!=\!3\pi/2$. The exact Floquet quasienergy spectrum obtained from the diagonalization of $\hat H_{\textrm{eff}}$ is shown in Fig.~\ref{fig2}(b) as a function of $K/\omega$ for the caging limit, $\theta\!=\!\pi/2$. Fig.~\ref{fig2}(c) presents a comparison of the exact quasienergy spectrum for the experimentally realized drive frequency (see below), $\omega/J\!=\!12.64$, and the one obtained from $\hat H_{\textrm{eff}}^{(0)}$ in Eq.~\eqref{Heff}, thus indicating that the high-frequency approximation is valid for our experiments.

In the AB caging limit ($\Phi\!=\!\pi$), the spectrum displays three non-dispersive bulk bands, Fig.~\ref{fig2}(b)-(c). One band is at energy $\epsilon_{0}^{\text{bulk}}\!=\!0$ and the localized eigenstates read $|\psi_{s,0}^{\text{bulk}}\rangle \!=\! (\hat b^\dagger_{s-1}+\hat c^\dagger_{s-1} + \hat b^\dagger_s - \hat c^\dagger_s )|0\rangle$~\cite{Note2}. The other two bands appear at energies $\epsilon_{\pm}^{\text{bulk}}\!=\!\pm 2 |J_{\textrm{eff}}|$ and the eigenstates are $|\psi_{s,\pm}^{\text{bulk}}\rangle\!=\! (\hat b^\dagger_{s-1}+ \hat c^\dagger_{s-1} \mp 2\hat a^\dagger_s - \hat b^\dagger_s + \hat c^\dagger_s )|0\rangle$. In Fig.~\ref{fig2}(b), we also see the appearance of states that are localized at the edges (highlighted by red lines). Indeed, when open boundary conditions are considered, such that the rhombic chain has an edge with an $A$ site termination, the effective model (\ref{Heff}) predicts a pair of states with energies $\epsilon_{\pm}^{\text{edge}}\!=\!\pm \sqrt 2 |J_{\textrm{eff}}|$; for the left edge of the chain, the eigenstates read $|\psi_{1,\pm}^{\text{edge}}\rangle\!=\!(\mp \sqrt 2 a^\dagger_1 - b^\dagger_{1}+c^\dagger_{1} )|0\rangle$. Similarly, a pair of edge states would appear on the right edge~\footnote{The small energy asymmetry observable in Fig.~\ref{fig2}(b-c) and the additional edge states at $\epsilon/J \approx 0$ owe their origin to low-frequency corrections in the effective model (\ref{Heff}), which we disregard in the rest of the discussion}.

Interestingly, the edge states of the rhombic chain with $\pi$-flux per plaquette can be continuously connected to the topological mid-gap edge state of the Creutz ladder~\cite{Creutz1999,Bermudez2017}. This can be seen by adding an onsite offset $\hat H_{\delta_A}\!=\!\delta_A \sum_s \hat a^\dagger_s \hat a^{}_s$ and by considering the limit $\delta_A \gg J_{\textrm{eff}}$; see Fig.~\ref{fig2}(d) and~\cite{Supplementary} for more details on this mapping.

{\it Experiments.$-$}
Our experimental platform consists of photonic lattices -- periodic arrays of evanescently coupled optical waveguides -- fabricated using ultrafast laser inscription~\cite{Davis1996writing, Supplementary}. 
Finite rhombic lattices with twelve unit cells were fabricated with a waveguide-to-waveguide separation $a/\sqrt{2}\!=\!17 \  \mu$m.
In the scalar-paraxial approximation~\cite{Christodoulides2003, Longhi2009quantum, Szameit2010}, the evolution of the optical field along the propagation distance ($z$) of our photonic lattices is formally equivalent to the time-evolution of a single particle wavefunction obeying the discrete Schr\"odinger equation associated with the Hamiltonian $\hat H_0$ in Eq.~\eqref{H0}. 
By characterizing a set of directional couplers (two identical evanescently-coupled waveguides) at $780$ nm wavelength, we obtained the analogous tunneling strength, $J\!=\!0.035\pm0.002 \ $mm$^{-1}$ entering Eq.~\eqref{H0}. The next-nearest-neighbor and higher-order tunneling processes were insignificant for the maximum propagation distance considered in this work. Within the mapping described above, the propagation constants of the optical modes play the role of onsite energies. Hence, the Hamiltonian $\hat H_\textrm d$ in Eq.~\eqref{H_dc} can be simulated by applying a linear gradient to the propagation constants; here this is obtained by circularly curving the axes of the waveguides in the photonic lattice~\cite{Lenz1999, Chiodo2006imaging, Mukherjee2015WSL}. Specifically, a radius of curvature $R$ leads to an energy shift $\Delta\!=\!n_0a/({2}R\lambdabar)$ in Eq.~\eqref{H_dc}, where $n_0$ is the average refractive index of the substrate and $\lambda\!=\!2\pi\lambdabar$ is the free-space wavelength of light. The driving term $\hat H_\textrm{ac}$ in Eq.~\eqref{H_ac} is then mimicked by applying a spatial square-wave modulation of the propagation constants, which is realized by varying the translation speed of fabrication; see Fig.~\ref{fig1}(d)  and~\cite{Supplementary}. In our experiment, this modulation has a period $z_0\!=\!2\pi/\omega\!=\!14 \ $mm, and it does not significantly affect the inter-site tunneling strength~\cite{Szameit2009inhibition, Mukherjee2016pair}. 
The peak-to-peak amplitude of the square wave modulation ($K$) was measured by characterizing a set of straight directional couplers fabricated with modulated translation speeds [$v_{1}\!=\!(v_0/2)f(\omega z)$ and $v_{2}\!=\!(v_0/2)f(\omega z+\pi)$]; see~\cite{Supplementary}.
The final modulated lattices with circularly-curved paths were fabricated inside a $70$-mm-long borosilicate substrate using two sets of extrema of translation speeds, $\{v_{\text{max}}, v_{\text{min}}\}\!=\!\{9, 6 \}$~mm/s and $\{9, 7 \}$~mm/s, which leads to the parameters values $K/\omega\!=\!1.35$ and $0.85$, respectively [see Eq.~\eqref{H_ac}]. Besides, in all the experiments described below, the resonance condition was set to $\Delta\!=\!\omega\!=\!12.64 J$.

\begin{figure}[t!]
\center
\includegraphics[width=8.6 cm]{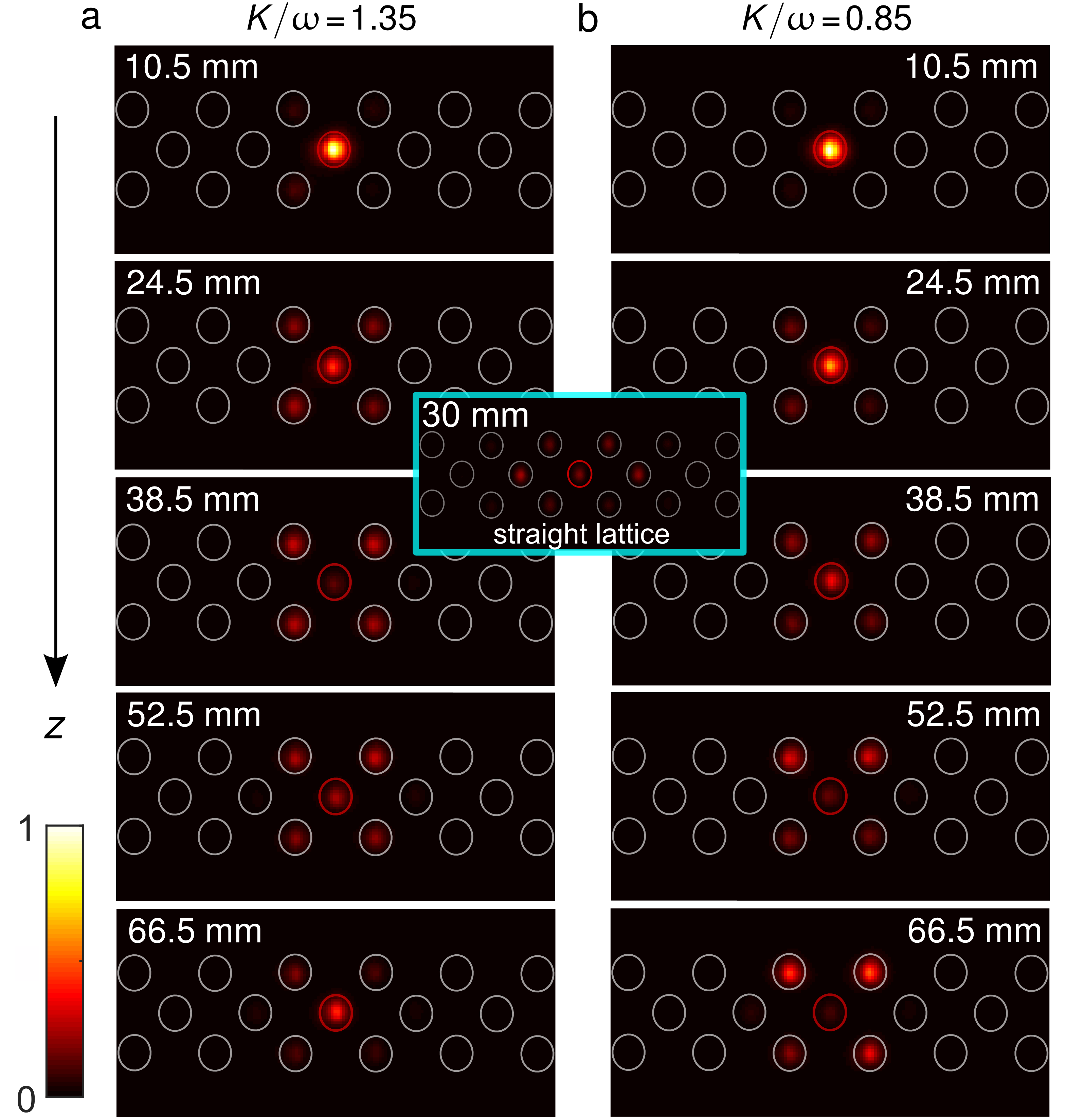}
\caption{Observation of Aharonov-Bohm photonic caging in the presence of $\Phi\!=\!\pi$ magnetic flux per plaquette. Experimentally measured output intensity distributions at five different propagation distances (shown on each image) for $K/\omega\!=\! 1.35$ (a) and $K/\omega\!=\! 0.85$ (b). The optical intensity exhibits a breathing motion whose frequency is determined by $K/\omega$. For all measurements, light was injected at the red circled $A$ site -- this input state overlaps efficiently with the upper and lower flat-bands and the frequency of the observed breathing motion is related to the energy gap between these two bands. Each image is normalized so that the total intensity is $1$. The field of view is approximately $121\ \mu{\text m} \times 61\ \mu{\text m}$. The inset shows the spreading (delocalization) of the optical intensity in a straight lattice (i.e.~$\Phi\!=\!0$); see~\cite{Supplementary}.
}
\label{fig3}
\end{figure}

The evolution of the optical intensity governed by the static Hamiltonian in Eq.~\eqref{H0}, in the absence of driving ($K\!=\!0$ and $\Delta\!=\!0$), is presented in~\cite{Supplementary}; see also the inset in Fig.~\ref{fig3}.
As shown in Fig.~\ref{fig2}(a), the synthetic magnetic flux can be activated in a modulated lattice, where it can be tuned by varying the phase of modulation. In~\cite{Supplementary}, we have presented the evolution of an input state, localized on a bulk $A$ site, with phases of modulation $\theta\!=\!\pi/5$ and $\pi$. Here in the main text, we focus on a specific value of flux, $\Phi\!=\!\pi$, for which optical waves tunneling along a closed loop on the lattice acquire a non-vanishing AB-type phase of $\pi$, thus realizing the photonic AB-caging regime. 
In our experiments, the maximum achievable strength of onsite modulation, which does not affect the tunneling strength, was found to be $K\!=\!1.35\,\omega$. In this situation, $\Phi\!=\!\pi$ magnetic flux per plaquette is realized for a phase $\theta\!=\!\pi/2$ and the effective tunneling strength becomes $|J_{\text{eff}}|\!=\!0.509J$; see Fig.~\ref{fig2}(a).

\begin{figure}[t!]
\center
\includegraphics[width=8.6 cm]{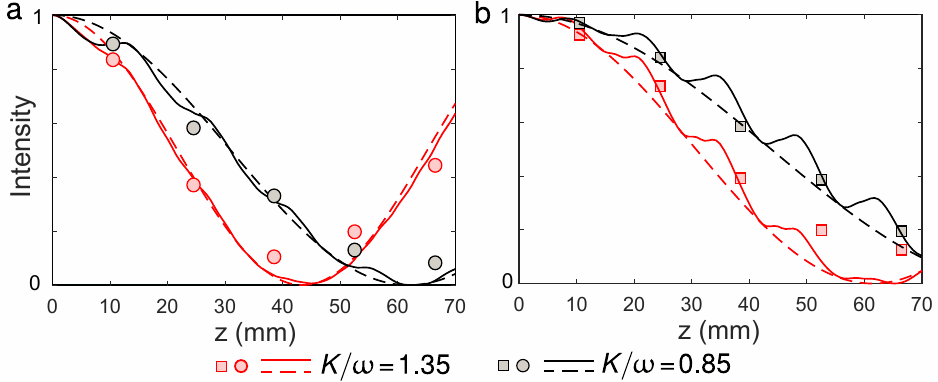}
\caption{(a) Graphical representation of Fig.~\ref{fig3}. The dashed and solid lines indicate the $z$ evolution of optical intensities (at the initially excited bulk $A$ site), which were numerically calculated solving the Schr{\"o}dinger equation for $\omega/J\!=\!12.64$  associated with the Hamiltonians $\hat H_{\textrm{eff}}$ and $\hat H(t\leftrightarrow z)$, respectively. Small oscillations can be observed as a consequence of micromotion. The filled circles indicate the measured light intensities at the $A$ site. See also Fig.~A3 in~\cite{Supplementary}. (b) Edge dynamics -- when light is launched at the $A$ site on the edge, the input state overlaps with edge modes located in both band gaps. In this case, a breathing motion with relatively lower frequency (compared to the bulk excitation) was observed, as expected. Indeed, the frequency of this breathing motion is associated with the energy difference between the edge modes [see Fig.~\ref{fig2}(c)].
}
\label{fig4}
\end{figure}

By launching light at $780 \ $nm wavelength into a bulk $A$ site, we excite a superposition of the upper and lower flat-band states and observe a breathing motion of the optical intensity, as shown in Fig.~\ref{fig3}(a). In this case, the expected period of oscillation, determined by the energy gap between the upper and lower flat-bands, is $\pi/(2|J_{\text{eff}}|)\!\approx\!88\ $mm. The optical intensity is trapped at the initially excited $A$ site and its four nearest neighbor sites due to the AB caging phenomenon. To demonstrate the tunability of the overall bandwidth of the spectrum, we perform another set of experiments with $K/\omega\!=\! 0.85$ and $\theta\!=\!\pi/2$, implying $J_{\text{eff}}/J\!=\!0.357$, see Fig.~\ref{fig3}(b). In this case, the frequency of the breathing motion is relatively lower, which reflects the smaller bandwidth of the spectrum; specifically, the expected period of oscillation is $\pi/(2|J_{\text{eff}}|)\!\approx\!124\ $mm. Fig.~\ref{fig4}(a) shows the comparison between the measured and expected variation of the optical intensities at the initially excited bulk $A$ site -- here, the dashed and solid lines indicate the $z$ evolution of the optical intensities which were numerically calculated for $\omega/J\!=\!12.64$ by solving the Schr{\"o}dinger equation associated with the Hamiltonians $\hat H_{\textrm{eff}}$ and $\hat H(t\leftrightarrow z)$, respectively. In the experiment, the oscillation-frequency was estimated from the half of the total period of oscillation, i.e.~the propagation distance at which the intensity at the initially excited $A$ site becomes minimal ($\approx 0$).
To further confirm the existence of three flat bands (AB caging), we launched light into a single bulk $C$ site, which has non-zero overlap with all three bands. In that case, the observed oscillations clearly revealed two frequencies, in agreement with the two relevant energy separations of the spectrum in Fig.~\ref{fig2}(c); see~\cite{Supplementary}.

In the next step, we launch light at the $A$ site on the edge -- this input state efficiently overlaps with the edge modes located in both band gaps. In the experiment, the optical intensity was observed to oscillate among the $A$ site (where light was initially launched) and its two neighboring $B$ and $C$ sites. As the energy gap between these edge modes is smaller compared to that between the upper and lower flat-bands [Fig.~\ref{fig2}(c)], we observe a breathing motion of optical intensity with relatively lower frequency (compared to the bulk $A$ site excitation), see Fig.~\ref{fig4}(b).

{\it Conclusion.$-$} 
We have experimentally demonstrated the realization of a uniform synthetic magnetic flux in ultrafast-laser-fabricated rhombic lattices.
This driving protocol can be extended to realize a uniform flux in 2D lattices, such as the square lattice, indicating an exciting route towards the experimental realization of the Hofstadter spectrum~\cite{Hofstadter_ref} and associated topological phenomena~\cite{Bernevig} in photonic lattices. 
In addition, the experimental platform offered by photonic lattices allows one to address and control specific lattice sites independently, which enables the realization and manipulation of magnetic flux on each individual plaquette; this makes the creation of spatially periodic or even random flux configurations~\cite{Bending1990weak, Mancoff1996shubnikov} accessible in experiments.

Realizing synthetic magnetic flux for light has allowed us to observe AB caging in photonic lattices, an effect that  originates from the isolated flat bands of the underlying spectrum. This is to be contrasted with other models that were previously implemented in the absence of magnetic flux, e.g.~the static rhombic lattice~\cite{Mukherjee2015rhombic} or the Lieb lattice~\cite{Mukherjee2015observation, Vicencio2015observation}, where flat bands always appear together with dispersive ones. We stress that these previous setups required specific state preparations to observe localization. Interestingly, in the case of AB caging, the role of interactions is enormously enhanced \cite{Douccot2002pairing,Rizzi2006,Huber2018}, even at the mean-field level, and it can lead to non-trivial states of matter where, for instance, time-reversal symmetry is spontaneously broken~\cite{Moller2012correlated}. This possibility therefore paves the way to investigate the impact of optical nonlinearities in fully gapped flat-band systems~\cite{Ozawa2018,Leykam2018a}.\\

Raw experimental data are available~\cite{data}.\\

{\it Acknowledgments.$-$} 
We thank M. Aidelsburger, E. Andersson, E. Anisimovas, F. Gerbier, A. Spracklen and M. Valiente for helpful discussions. N.G.~and M.D.~acknowledge support from the ERC Starting Grant TopoCold. This work was funded as part of the UK Quantum Technology Hub for Quantum Communications Technologies - EPSRC grant no. EP/M013472/1, and by the UK Science and Technology Facilities Council (STFC) - STFC grant no. ST/N000625/1. S.M. thanks Universit\'e Libre de Bruxelles (ULB) and Scottish Universities Physics Alliance (SUPA) for hosting and funding through the Postgraduate, Postdoctoral and Early Career Researcher Short-Term Visits Programme-2018, respectively. P.\"O.~acknowledges support from EPSRC grant No. EP/M024636/1.


{\it Note added.$-$} After completion of this work, we were made aware of similar results reported in Ref.~\cite{Szameit2018}.
%

\newcommand{\beginsupplement}{%
        \setcounter{equation}{0}
        \renewcommand{\theequation}{A\arabic{equation}}%
        \setcounter{figure}{0}
        \renewcommand{\thefigure}{A\arabic{figure}}%
     }
\beginsupplement

\section*{Supplemental Material}
{\it Fabrication and characterization details.$-$}
Here, we discuss fabrication and characterization of photonic devices for the purpose of completeness. First, we briefly present the experimental parameters for fabricating optical waveguides and then, discuss how the square-wave modulation of onsite energy was realized.

Circularly polarized sub-picosecond ($350 \ $fs) optical pulse trains of $500 \ $kHz repetition rate and $1030 \ $nm wavelength, generated by a Yb-doped fiber laser (Menlo Systems, BlueCut), were used to fabricate optical waveguides inside borosilicate glass (Corning Eagle$^{2000}$) substrates. Each waveguide was inscribed by translating the substrate once through the focus of the laser beam. The optical pulse energy of the laser was optimized to inscribe tightly confined singlemode optical waveguides at $780$~nm wavelength and for a range of translation speed of fabrication -- i.e.~$9 \ $mm/s to $5 \ $mm/s. Inter-site tunneling strength was obtained by measuring light intensities at the output of straight directional couplers (two coupled waveguides) fabricated with the same waveguide-to-waveguide separation (i.e.~$17 \ \mu$m) and waveguide-to-waveguide angle as the waveguides in the rhombic lattice. 

\begin{figure}[]
\center
\includegraphics[width=8.6 cm]{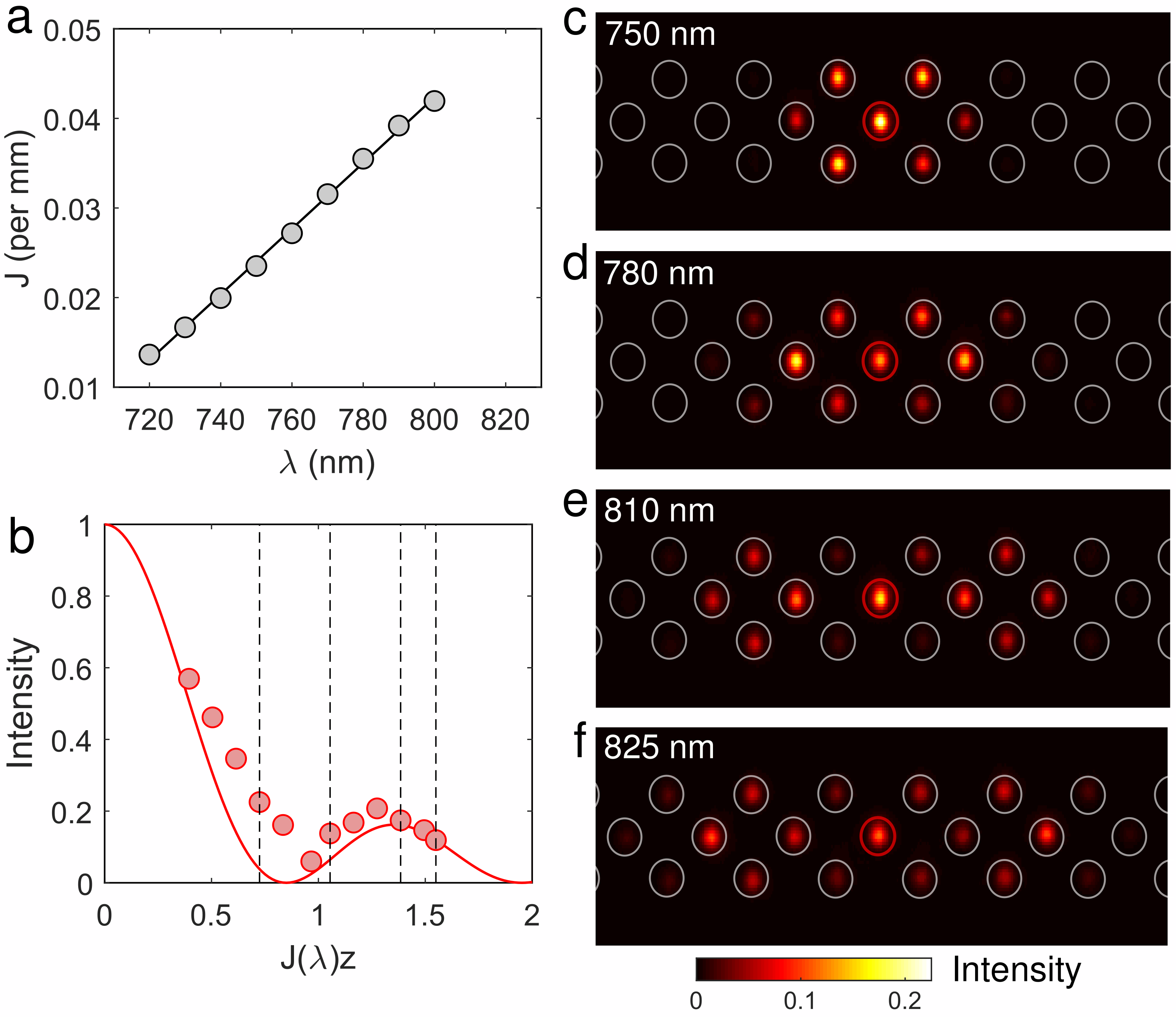}
\caption{(a) Variation of inter-site tunneling strength, $J$, as a function of the wavelength of incident light, $\lambda$. (b) Evolution of light intensity at the bulk $A$ site (where the light was launched initially) in a $30 \ $mm long straight photonic rhombic lattice. Instead of measuring light intensity at different $z$ values with a fixed $J$, the wavelength of incident light was tuned to vary the inter-site tunneling strength and the intensity distributions were measured at $z\!=\!30 \ $mm. (c-d) Measured output intensity distributions for four different values of $J(\lambda)z$, indicated by the vertical dotted lines in (b). For all measurements, the light was launched at the red-circled central $A$ site. Each image is normalized so that the total intensity is $1$. The field of view is approximately $162\ \mu{\text m} \times 61\ \mu{\text m}$.
}
\label{supp_1}
\end{figure}

Fig.~\ref{supp_1}(a) shows the variation of tunneling strength, $J$, as a function of the wavelength of incident light -- note that $J$ varies linearly with $\lambda$. To observe the evolution of the optical intensity, which is governed by the static Hamiltonian  Eq.~(1) 
(i.e.~in the absence of driving, namely~$K\!=\!0$ and $\Delta\!=\!0$), a $30$~mm-long straight lattice was fabricated with $17 \ \mu$m inter-waveguide spacing and $6 \ $mm/s translation speed of fabrication.  
For a given initial state, the dynamics of intensity pattern along this straight photonic lattice is determined by the parameter $Jz$.  
Instead of measuring light intensity at different $z$ values with a fixed $J$, the wavelength of incident light was tuned to vary inter-site tunneling strength and the intensity distributions were measured at $z\!=\!30 \ $mm. Figs.~\ref{supp_1}(b-f) show the measured dynamics. It should be mentioned that the wavelength tuning was performed only for this particular experiment -- measuring intensity evolution in the static case. All the experiments, detailed in the main text and also in the following sections of the supplementary material,
were performed at $780 \ $nm wavelength because the energy shift due to the static field, $\Delta\!=\!n_0a/({2}R\lambdabar)$, depends on $\lambda$ and the wavelength tuning would change the resonance condition, $\Delta\!=\!\omega$, for the driven lattices.

\begin{figure}[]
\center
\includegraphics[width=8.6 cm]{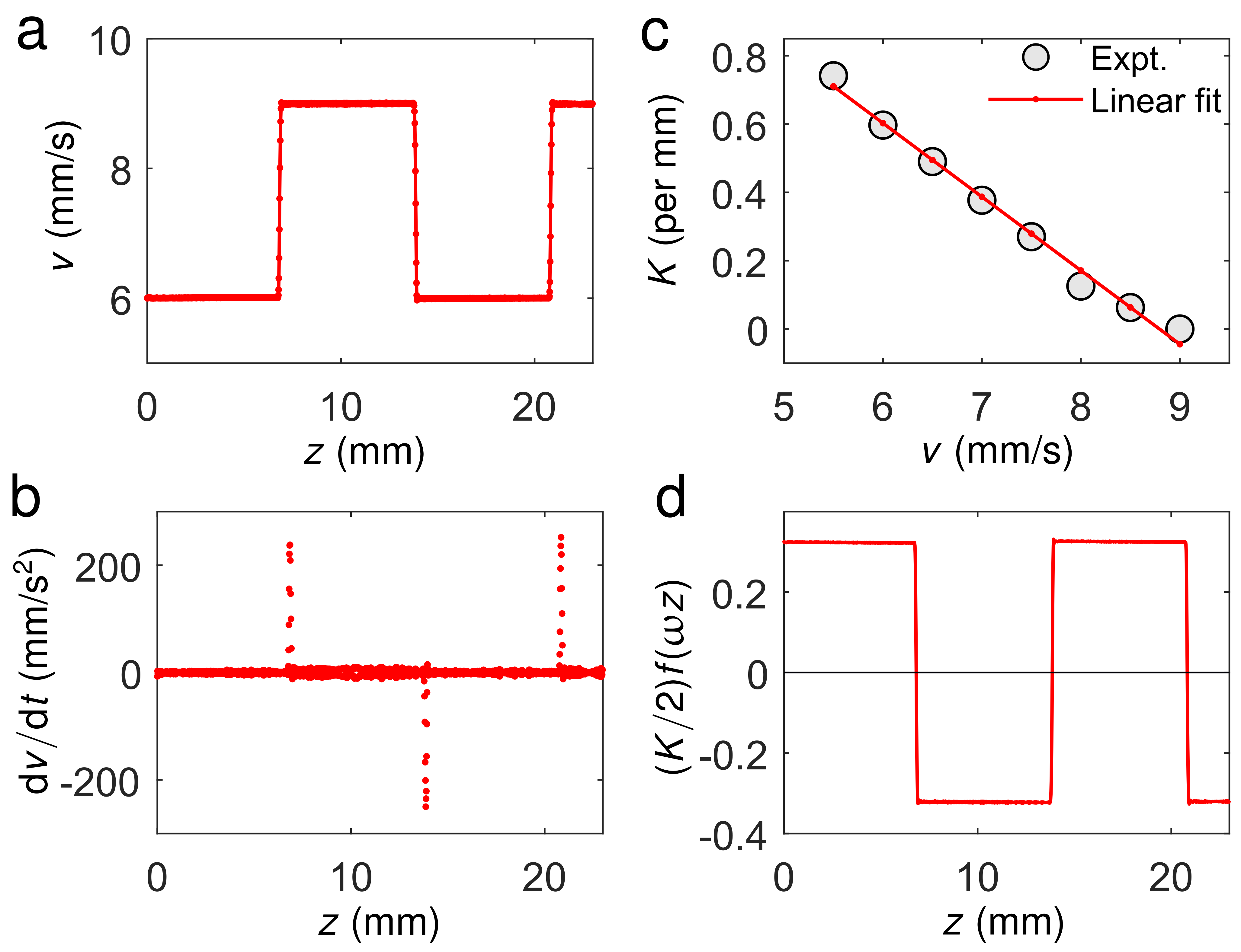}
\caption{(a) Measured variation of translation speed, $v$, along the propagation distance, $z$. (b) Measured variation of acceleration of the translation stages associated with (a). (c) Variation of the shift in propagation constant as a function of $v$. Here, $K(v\!=\!9 {\text{mm/s}})$ was chosen as the reference. (d) The modulation of propagation constant, $(K/2)f(z)$, which was estimated using the calibration in (c). Here, the average of this periodic modulation over a complete cycle is chosen to be the reference ($=0$).
}
\label{supp_2}
\end{figure}

\begin{figure}[]
\center
\includegraphics[width=8.6 cm]{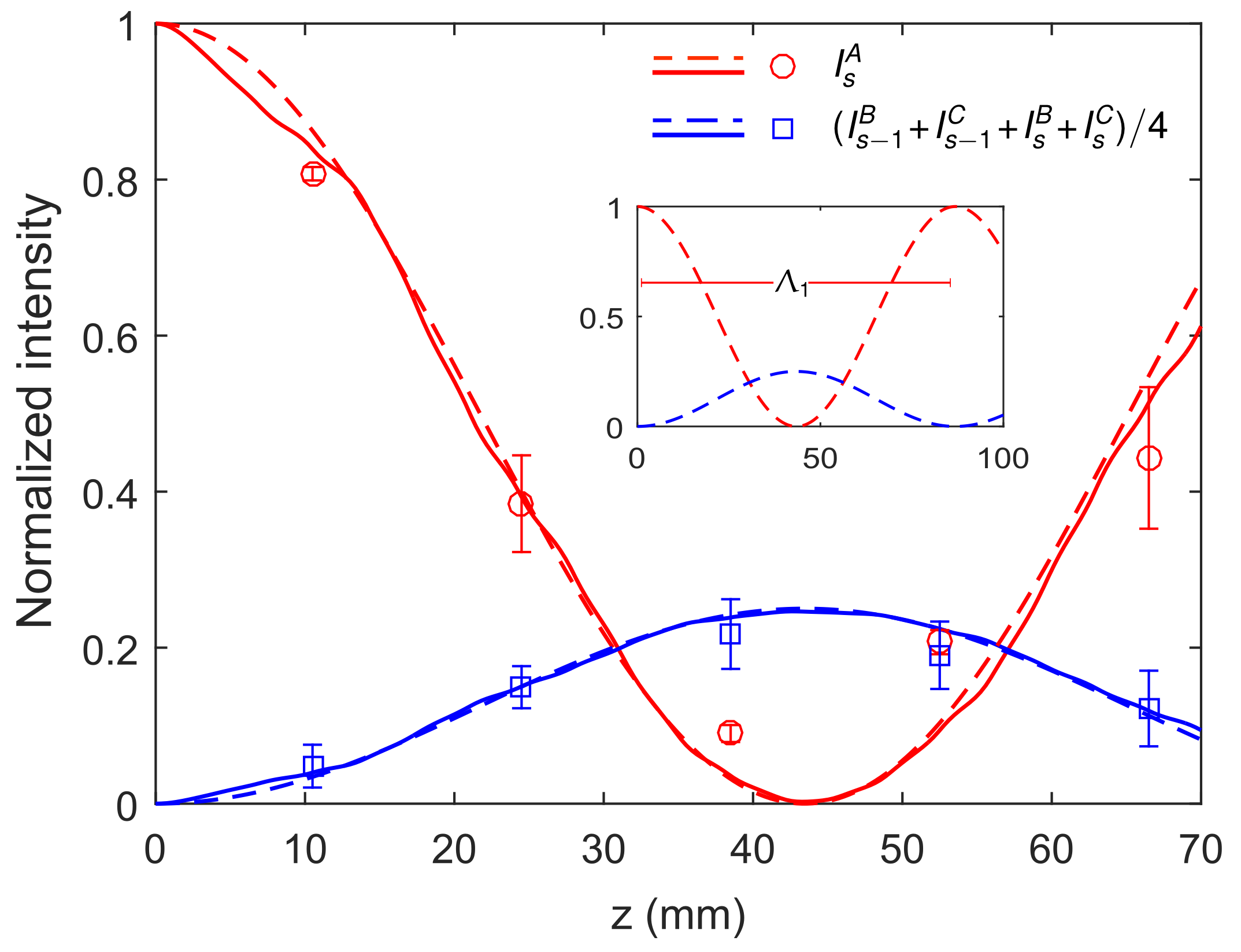}
\caption{Observation of AB caging when light was launched into a single bulk $A$ site -- graphical representation of Fig. 3(a) in the main text. This input state overlaps with the upper and lower flat-bands and the optical intensity oscillates with a period $\Lambda_1$ associated with the band gap ($2\epsilon$) between the bands. The solid and dashed lines were obtained considering the Hamiltonians, $\hat H(t\!\leftrightarrow\!z)$ and $\hat H_{\text{eff}}$ [see main text]. Every data point is an average over three individual measurements performed by launching light into $A$ sites of three different unit cells. The error bars show the standard deviation of our data. The inset shows the full dynamics.
}
\label{Asite}
\end{figure}

Using ultrafast laser inscription, we tune the refractive index contrast (the difference between the refractive index of the core and that of the substrate) of an optical waveguide by varying the translation speed ($v$) of fabrication and keeping all other fabrication parameters (including the optical pulse energy) unaltered. To realize a square wave modulation of the propagation constant (the analogous onsite energy) along the propagation distance, the translation speed is modulated periodically, as shown in Fig.~\ref{supp_2}(a). Fig.~\ref{supp_2}(b) presents the variation of acceleration of the translation stages along the propagation distance. We calibrate the shift in propagation constant as a function of $v$ in the following way.
A set of straight directional couplers (two coupled waveguides) were fabricated with modulated translation speeds, $v_{1}\!=\!(v_0/2)f(\omega z)$ and $v_{2}\!=\!(v_0/2)f(\omega z+\pi)$. By measuring the light intensities at the output of these modulated couplers, the modulus of the effective tunneling is obtained as a function of $v_0$.  This experimental dependence of the effective tunneling is then fitted with its calculated variation with $K$, which gives a calibrated variation of $K$ with the translation speed of fabrication, see Fig.~\ref{supp_2}(c). Using this calibration, the modulation of propagation constant, $(K/2) f(z)$, was estimated as shown in Fig.~\ref{supp_2}(d). 
We also measured the variation of inter-site tunneling, $J (\lambda\!=\!780 \ {\text{nm}})$, as a function of $v$~[42, 43] 
and it was observed that $J$ does not vary significantly from $9 \ $mm/s to $6 \ $mm/s range of translation speed.
In other words, the maximum achievable strength of onsite modulation, without affecting the tunneling strength, was $K\!=\!1.35\,\omega$ (for $2\pi/\omega\!=\!14 \ $mm).

\begin{figure}[t]
\center
\includegraphics[width=8.6 cm]{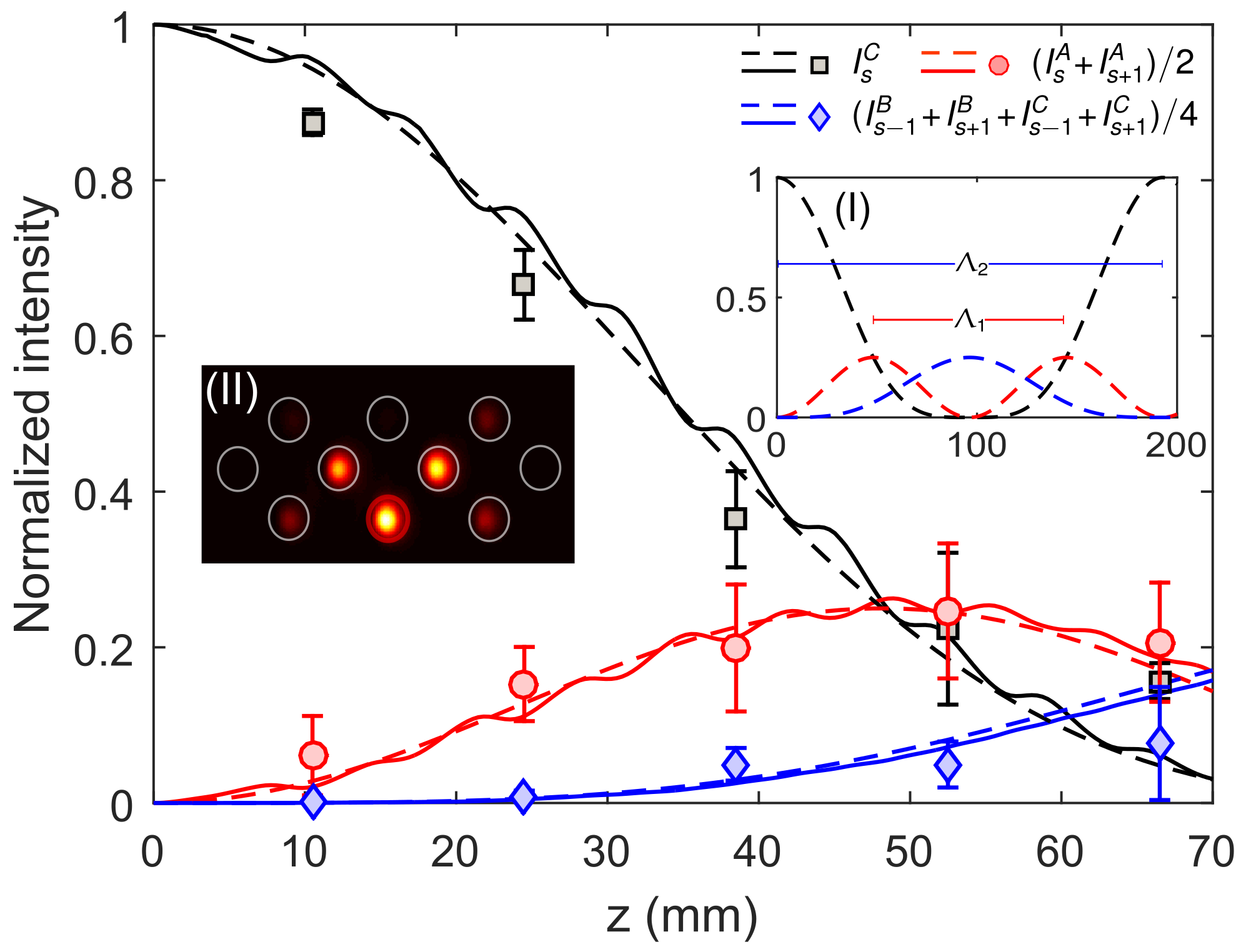}
\caption{Observation of AB caging when light was launched into a single bulk $C$ site. This input state overlaps with all the three flat-bands and the breathing motion of optical intensity exhibits two different periods of oscillation, $\Lambda_1$ and $\Lambda_2$ associated with the two band gaps ($2\epsilon$ and $\epsilon$); see inset (I) for the complete dynamics. Every data point is an average over three individual measurements performed by launching light into $C$ sites of three different unit cells. The error bars show the standard deviation of our data. Inset (II) shows an example of measured intensity pattern at $z\!=\!38.5 \ $mm. In this set of experiments, $K/\omega\!=\!1.35$ and the inter-site tunneling strength $J\!=\!0.032$~mm$^{-1}$.}
\label{Csite}
\end{figure}

{\it More details on AB caging.$-$}
In the main text, we have presented experimental results on AB caging when light was launched into a single bulk $A$ site; see Fig.~3(a, b) and Fig.~4(a). This input state [$|\psi(t\!=\!0)\rangle \!= \!\hat a^\dagger_s |0\rangle\!\equiv\!|A_s\rangle$] overlaps with the upper and lower flat-bands at quasienergies $\epsilon_\pm\!=\!\pm2 |J_\textrm{eff}|\equiv \pm\epsilon$, respectively (the middle band is at $\epsilon_0\!=\!0$). Considering the effective model, the state at time $t$ can be written as $|\psi(t)\rangle\!=\! \exp(-i \hat H^{(0)}_{\textrm{eff}}t)|\psi(t\!=\!0)\rangle$, where the micromotion is neglected and therefore, the time evolution is quantitatively correct only at stroboscopic times. We can extract the probability to find the particle on the $A_s$, and $B_{s}$ (and/or $B_{s-1}$, $C_{s}$ \& $C_{s-1}$) sites (i.e.~the analogous optical intensities at these sites) as
\begin{eqnarray}
I_s^A \!&=&\! |\langle A_s|\psi(t)\rangle|^2\!=\! \frac{1}{2} \big(1+\cos(2\epsilon t) \big) \\
I_s^B \!&=&\!|\langle B_s|\psi(t)\rangle|^2\!=\! \frac{1}{8} \big(1-\cos(2\epsilon t) \big)  \\
&=&\!I_s^C \!=\!I_{s-1}^B \!=\!I_{s-1}^C 
\end{eqnarray}
Note that the optical intensity at the initially excited bulk $A$ site (i.e.~$I_s^A$) and that at the four nearest $B$ and $C$ sites oscillate with a period associated with the band gap between the upper and lower bands, i.e.~$\Lambda_1\!=\!2\pi/(2\epsilon)\!=\!\pi/(2 |J_\textrm{eff})|$. In Fig.~4(a), we have presented the oscillation of $I_s^A$ . Supplementary Fig.~\ref{Asite} shows the complete information i.e.~the dynamics of $I_s^A$ as well as its four nearest neighbors. Every data point in Fig.~\ref{Asite} is an average over three individual measurements performed by launching light into $A$ sites of three different unit cells and the error bars show the standard deviation of our data. 

We have performed another set of experiments to further confirm the observation of Aharonov-Bohm caging and the corresponding fully flat spectrum by launching light into a bulk $C$ site. In this situation, differently from the previous experiments presented in Fig.~3(a, b), Fig.~4(a) and Fig~\ref{Asite}, a superposition with all three bands occurs. 
In this case, the optical intensity at $C_s$, $A_s$ (and/or $A_{s+1}$) and $B_{s+1}$ (and/or $B_{s-1}$, $C_{s\pm1}$) sites can be written as
\begin{align}
&I_s^C \!=\! |\langle C_s|\psi(t)\rangle|^2\!=\!\frac{1}{8} \big(3+\cos (2\epsilon t) + 4 \cos (\epsilon t)\big)\\
&I_s^A \!=\!|\langle A_s|\psi(t)\rangle|^2\!=\!\frac{1}{8} \big(1-\cos (2 \epsilon t)\big)\!=\! I_{s+1}^A\\
&I_{s+1}^B\!=\!\frac{1}{32} \big(3+\cos (2\epsilon t) - 4 \cos (\epsilon t)\big)\!=\! I_{s-1}^B\!=\! I_{s\pm1}^C
\end{align}
The breathing motion of optical intensity exhibits  two different dominant harmonics -- $I_s^A$ oscillates with period 
$\Lambda_1\!=\!\pi/\epsilon$ whereas $I_s^C$ (and/or $I_{s+1}^B$) oscillates with a period twice longer, namely $\Lambda_2\!=\!2\pi/\epsilon$. 
Notice that the expression for $I_C$ contains two frequencies of oscillation. The strengths of the two harmonics are such that the signal displays no evident beating effect but a flattening of the signal at $t = \pi/\epsilon$, corresponding to half a period of the dominant harmonic. The agreement of the experiments and theory is shown in Fig.~\ref{Csite}.

\begin{figure}[]
\center
\includegraphics[width=8.6 cm]{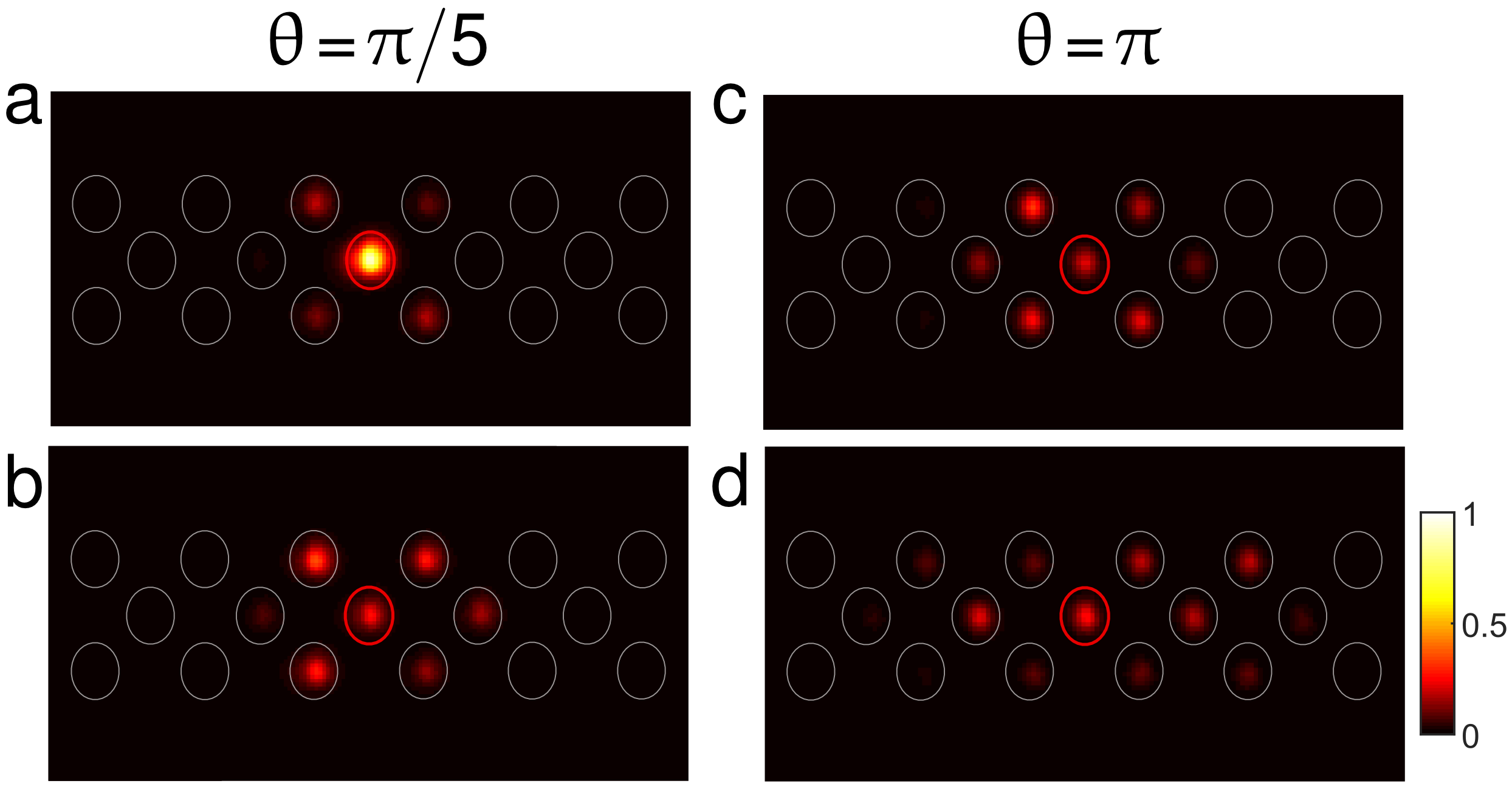}
\caption{(a, b) Experimentally measured output intensity distributions at $z\!=\!35 \ $mm and $70 \ $mm, respectively, for $\theta\!=\! \pi/5$. The red circled $A$ site was initially excited for all measurements. Here, $K/\omega\!=\!1.35$. (c, d) Same as (a, b) with $\theta\!=\! \pi$.}
\label{otherFlux}
\end{figure}

{\it Realization of $2\pi$ and $8\pi/5$ flux per plaquette.$-$}
So far, we have mainly focused on the experimental demonstration of AB caging in the presence of $\pi$ magnetic flux per plaquette. As shown in Fig.~2(a), any desired value of uniform magnetic flux can be realized by correctly choosing the phase of modulation, $\theta$. To demonstrate this experimentally, we fabricated two sets of circularly curved modulated lattices with  $K/\omega\!=\!1.35$ and the phases of modulation $\theta\!=\!\pi/5$ and $\pi$, respectively . For these two sets of experiments, $|J_{\text{eff}}|/J\!=\!0.258, 0.546 \ $ and  $\Phi\!=\!8\pi/5, 2\pi$, respectively. Fig.~\ref{otherFlux} shows experimentally measured output intensity distributions at $z\!=\!35$ and $70 \ $mm.

{\it Details on the Floquet theory.$-$}
To construct an effective description of the driven system described by the time-dependent Hamiltonian $\hat H(t)\!=\!\hat{H}_{0}+\hat{H}_{\text{d}}+\hat{H}_{\textrm{ac}}$, we consider a unitary transformation $\hat{\mathcal R}$ such that the Hamiltonian transforms as $\hat H'(t)\!=\!\hat{\mathcal R}\hat H(t)\hat{\mathcal R}^\dagger-i\hbar \hat{\mathcal R}\partial_t \hat{\mathcal R}^\dagger$. Following the standard approach for resonant modulations~[24, 32, 33] 
and taking $\Delta\!=\!\omega$, we define the unitary transformation $\hat{\mathcal R}\!=\!\hat{\mathcal R}_A\hat{\mathcal R}_B\hat{\mathcal R}_C$ where
\begin{align}
\hat{\mathcal R}_A &\!=\!\exp\Big[ i \sum_s\! \big(\!- \!2s \omega t + \frac{K}{2\omega} \chi(\omega t) \big) \hat{a}^{\dagger}_s \hat{a}_s \Big],\nonumber \\
\hat{\mathcal R}_B &\!=\!\exp\Big[ i \sum_s\! \big(\!- \!(2s+1) \omega t + \frac{K}{2\omega} \chi(\omega t+\theta) \big) \hat{b}^{\dagger}_s \hat{b}_s \Big], \nonumber\\
\hat{\mathcal R}_C &\!=\! \exp\Big[ i \sum_s\! \big(\!- \!(2s+1) \omega t + \frac{K}{2\omega} \chi(\omega t-\theta) \big) \hat{c}^{\dagger}_s \hat{c}_s \Big],
\end{align}
and $\chi(\omega t)\!=\!\omega\int^t f(\omega \tau)\textrm{d} \tau$. Differently from Ref.~[24] 
where the modulation was sinusoidal, in the experiment we are using a square wave function that can be conveniently written as 
\begin{equation}
f(\omega t)\!=\!-\sum_p \frac{4}{\pi(2p+1)}\sin[(\omega t - \pi)(2p+1)]\,,
\end{equation}
from which the function $\chi(\omega t)$ can be readily calculated. In the end, we consider the time average of the Hamiltonian $\hat H'(t)$, which corresponds to the lowest order in $1/\omega$ of the operator $\hat H_{\textrm{eff}}$, namely 
\begin{equation}
\hat H_{\textrm{eff}} = \frac 1 T \int_0^T \textrm d t \, \hat H'(t) + O(1/\omega) \,.
\end{equation}
The model takes the form (4) 
where, for instance, the restored hopping $J(B_s\rightarrow A_s)$ from a $B$ site to an $A$ site in the same unit cell $s$ is 
\begin{equation}
J(B_s\rightarrow A_s) = \frac 1 T \int_0^T \textrm d t\, e^{i\omega t} e^{iK(\chi(\omega t)-\chi(\omega t + \theta))/2\omega}\,.
\end{equation}
The parameters of the model (4) 
can then be calculated as $\theta_1\!=\!\textrm{arg} [ J(B_s\rightarrow A_s) ]$ and $|J_{\textrm{eff}}|\!=\!\textrm{abs} [J(B_s\rightarrow A_s)]$. Similarly, one obtains the parameters of all the bonds. While the phases $\theta_j$ are not necessarily going to be all the same depending on the choice of the unitary transformation $\hat{\mathcal R}$, notice that the total flux $\Phi\!=\!\sum_ j\theta_j$ will be a gauge invariant quantity.

{\it Mapping to the Creutz ladder.$-$}
In this section, we show how to map the model in Eq.~(4) 
which describes a single particle hopping in a rhombic lattice with non-vanishing flux $\Phi$, into the Creutz ladder~[26]. 
In particular, it is convenient to consider a gauge where each bond has the same hopping phase $\bar \theta\!=\!\theta_1\!=\!\theta_2\!=\!\theta_3\!=\!\theta_4$.

\begin{figure}[]
\center
\includegraphics[width=7.2 cm]{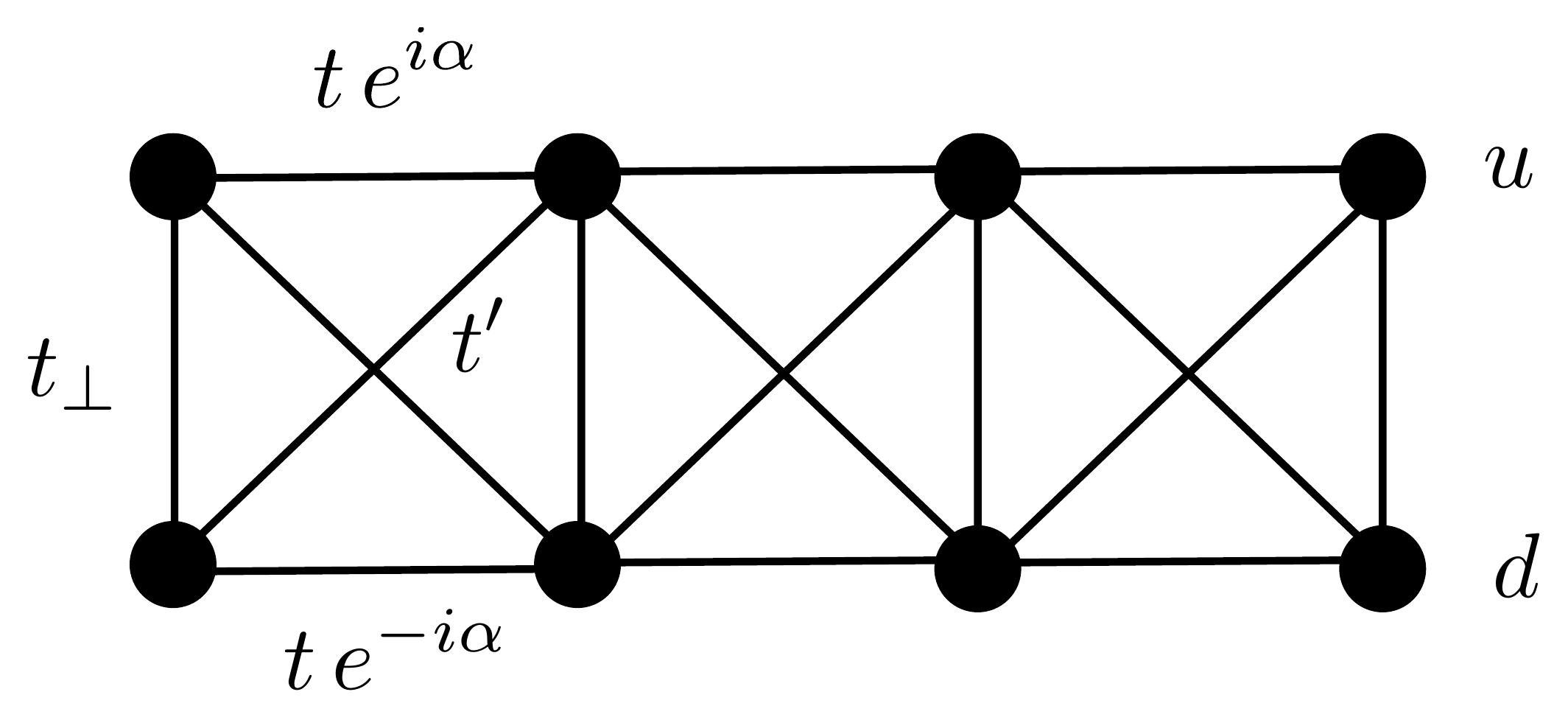}
\caption{Creutz ladder obtained as a low energy model for the rhombic lattice with a non-vanishing flux when the onsite energy of the $A$ sites is largely detuned.}
\label{Creutz}
\end{figure}

Let us add to the Hamiltonian (4) 
an energy off-set on the $A$ sites $\hat H_{\delta_A}=\delta_A\sum_s \hat{a}^{\dagger}_s \hat{a}^{}_s$ such that $\delta_A \gg J_{\textrm{eff}}$. The low-energy physics is limited to the dynamics on the $B$ and $C$ sites through virtual processes occurring via the intermediate $A$ sites. The effective geometry for the $B$ and $C$ dynamics now resembles a ladder, as shown in Fig.~\ref{Creutz}, where we re-labeled $B\rightarrow u$ and $C\rightarrow d$. The low-energy model reads
\begin{align}
\hat{H}_{\textrm{Creutz}}\!=\!& \, + t\sum_s (e^{i\alpha} \hat{u}^\dagger_{s+1} \hat{u}^{}_{s}+e^{-i\alpha} \hat{d}^\dagger_{s+1} \hat{d}^{}_{s}+\textrm{H.c}) \nonumber\\
&\, + t_\perp \sum_s (\hat{u}^\dagger_s \hat{d}^{}_s+\textrm{H.c})\nonumber \\
&\, + t' \sum_s(\hat{d}^\dagger_{s+1}\hat{u}^{}_s + \hat{u}^\dagger_{s+1}\hat{d}^{}_s+\textrm{H.c.})\nonumber\\
&\, +\mu\sum_s(\hat{u}^\dagger_s \hat{u}^{}_s + \hat{d}^\dagger_s \hat{d}^{}_s) \,,
\end{align}
where $t\!=\!-J_{\textrm{eff}}^2/\delta_A$, $t_\perp \!=\! -2\cos(2\bar \theta)J_{\textrm{eff}}^2/\delta_A$, $t'\!=\!t$, $\mu\!=\! -2J_{\textrm{eff}}^2/\delta_A$ and $\alpha\!=\!2\bar\theta$.

In momentum space, the Hamiltonian reads
\begin{equation}
H(k) = 
\begin{pmatrix}
\mu + 2t\cos(k+\alpha)  & t_\perp + 2 t'\cos k\\ 
t_\perp + 2 t'\cos k & \mu + 2t\cos(k-\alpha)
\end{pmatrix}\,,
\end{equation}
which can also be written as $H(k)\!=\!{\boldsymbol{d}}(k)\cdot {\boldsymbol{\sigma}} + d_0(k)$ where ${\boldsymbol{\sigma}}\!=\!\{\sigma_x,\sigma_y,\sigma_z \}$ are the Pauli matrices and 
\begin{align}
d_0(k) &= \mu + 2t\cos\alpha\cos k\,,\nonumber\\
d_x(k) &= t_\perp + 2 t' \cos k\,,\nonumber\\
d_y(k) &= 0\,,\nonumber\\
d_z(k) &= -2t \sin\alpha\sin k\,.
\end{align}
In the cage limit, which is realized when $\bar \theta \!=\! \pi/4$ and therefore $\alpha \!=\!\pi/2$ and $t_\perp\!=\!0$, we immediately see that $d_0(k) \!=\!\mu$ is just an energy shift. The spectrum is $E_{\pm}^{\text{Creutz}}\!=\!\mu\pm2|t|$ and therefore it has two flat-bands at energy $E_{+}^{\text{Creutz}}\!=\!0$ and $E_{-}^{\text{Creutz}}\!=\!-4J^2_{\textrm{eff}}/\delta_A$ (see red dashed line in Fig.~2(d), main text). 
In this limit, a chiral symmetry emerges $\{ \sigma_y, H(k)\}\!=\!0$, thus allowing the existence of mid-gap topological edge states protected by chiral symmetry and a non-vanishing winding number $w\!=\!1$~[27]. 

\end{document}